ORGANISATION EUROPÉENNE POUR LA RECHERCHE NUCLÉAIRE
**CERN** EUROPEAN ORGANIZATION FOR NUCLEAR RESEARCH

# Simon van der Meer
# and
# his legacy to CERN and particle accelerators

V. Chohan

# Abstract

Simon van der Meer was a brilliant scientist and a true giant in the field of accelerators. His seminal contributions to accelerator science are essential to this day in our quest to satisfy the demands of modern particle physics. Whether we are talking of long-baseline neutrino physics or antiproton-proton physics at CERN and Fermilab, or proton-proton physics at the LHC, his techniques and inventions have been a vital and necessary part of modern-day successes. Simon van der Meer and Carlo Rubbia were the first CERN scientists to become Nobel laureates in Physics in 1984. His less well-known contributions spanned a whole range of subjects in accelerator science from magnet design to power supply design, beam measurements, slow beam extraction, sophisticated programs, and controls.

# Preface

While writing a homage to Simon van der Meer for another review last year, I soon realized that I had to start with a caveat. He was a practical genius who was often unjustifiably feared, because he could seldom be equalled or challenged, and, invariably, he would be right in any scientific discourse. When approached with a novel suggestion, he would often have had the same idea already and knew whether or not it would work, thereby restraining discussion. One could easily get short shrift if one argued with him in a mistaken belief or if one disturbed him during his contemplative moments. Nevertheless, he was a kind and honest person with a very brilliant, fast-thinking brain. Many at CERN since the 1960s knew Simon's temperament, his shyness, and his perhaps taciturn nature. However, those who came to know him closely realized quickly that his modesty and kindness never excluded helping those who sought his counsel or help. So, although he did not suffer fools gladly, a fact acknowledged even by his children, Esther and Mathijs, he was indeed very practical in his approach, and was helpful to those acceptable to him, as well as being modest. The latter is best illustrated by his statement during a panel discussion on the future of particle physics chaired by Carlo Rubbia in September 2003, where he was among ten eminent personalities and Nobel Prize winners. This statement followed a day-long symposium on 'Prestigious discoveries at CERN, 1973 & 1983' [1]:

> Pierre [Darriulat] just said that he was embarrassed. I think that I have the right to be even more embarrassed because I feel a complete outsider in this company. I have never done any particle physics and even in the range of machine physics, I always felt like an amateur. So I will not say anything about the future of CERN, this is really beyond me. But what I want to say to all people who are working on machine physics is to think of two things: first of all, do not believe it when people tell you that something is impossible. Always try to follow up crazy ideas. And don't forget that all the experts in machine physics sometimes forget things which you can do by making some kind of 'bricolage', those things, which people thought could not work, still work, if you work on it long enough. I think that's all I want to say. Thank you.

To lend weight to my personal experience and belief, I cannot help but cite Martinus Veltmann's statement, just following Simon:

> I think that I have seldom seen such a spectacle of modesty as in Simon's view on these things. Few people have contributed so much to this Laboratory as he has. Thank you.

# Contents

# 1 Introduction

Simon's practical approach to all aspects of work and life and his versatility are legendary. This, coupled with his prolific inventiveness, meant that he was arguably the cleverest person that I have come across in my thirty-odd years at CERN. This statement also reflects the views of Hans-Karl Kuhn from his days with the Super Proton Synchrotron (SPS) power supplies. Simon's contributions to CERN and accelerator physics speak for themselves. These started with magnet design in the 28 GeV Proton Synchrotron (PS) era in the 1950s and the 1961 invention of a pulsed focusing device, known as the 'van der Meer horn'. This was followed in the 1960s by the design of a small storage ring for a physics experiment studying the anomalous magnetic moment of the muon. Soon after and in the following decade, he did some very innovative work on the regulation and control of power supplies for the Intersecting Storage Rings (ISR) and, later, the SPS. His ISR days in the 1970s led to his technique for luminosity calibration of colliding beams, first used at the ISR and still used today at the LHC, as in other colliders. Last, but not the least was the Nobel Prize-winning idea behind stochastic cooling and the application of that at CERN in the late 1970s and 1980s. Simon's prolific inventiveness means that the whole park of accelerators at CERN that run so well today for physics, whether they might be for neutrinos sent to Gran Sasso, colliding proton beams at the LHC, or antiproton physics at the Antiproton Decelerator (AD), owe him an immense amount of gratitude. Likewise, the Fermilab antiproton programme that has been running since 1983–85 on the other side of the Atlantic, and the successes of the $p$–$\bar{p}$ Tevatron Collider up to 2011 and its discovery of the top quark, owe him considerable gratitude.

# 2 Early years

During his formative years in Holland, he was known to have frequented the local flea markets, taking home his purchases to invent and repair all things electromechanical at his parents' home. When asked once what he was doing when the war broke out and everyone was glued to their radios listening, Simon's typically nonchalant answer was that he was probably repairing a radio instead! When refrigerators were still new in Holland and lacked a light that came on when the door was opened, he invented a light for the refrigerator at home that did exactly that.

Simon was born in The Hague in 1925, the third child of Pieter van der Meer and Jetske Groeneveld. His father was a schoolteacher and his mother came from a teacher's family, and hence it was hardly surprising that education and learning were highly cherished in the van der Meer family; indeed, his parents made sacrifices to educate Simon and his three sisters. Having attended high school (*gymnasium*) in The Hague, in the science section, he passed his final examination in 1943, during the German occupation of Holland during the war. Although he was qualified to pursue further education, he could not go to university because the German occupation had closed the universities. So he stayed on in high school, attending classes in the humanities section but at the same time assisting his ex-physics teacher with the preparation of numerous school physics demonstrations. This was a highly formative period for Simon; his interest in physics and technology knew no bounds and the physics teacher provided much inspiration and encouragement. Simon avidly dabbled in electronics, as he would say, equipping the family home with various gadgets, including the fridge light!

In 1945, Simon began studying ‘Technical Physics’ at Delft University, where he specialized in measurement and regulation technology. This was a precursor to his influential work on voltage-controlled power supplies and closed-loop control at the ISR and SPS. The technical-physics knowledge that Simon acquired at Delft was excellent, albeit “restricted out of necessity”, to use Simon’s words. He regretted not having the intensive physics training that his contemporaries had acquired; however, he felt that his substantial practical experience coupled with what he called a “slightly amateur” approach to physics were an asset that enabled him to make original contributions in the field of accelerators. In a similar vein to his decisive work on power supplies, his training in measurement and feedback at Delft also led him to his well-acknowledged invention, stochastic cooling, which is also a combination of measurement (of the position of particles) and feedback.

After obtaining his degree in 1952, he joined the Philips research laboratory in Eindhoven, working mainly on high-voltage equipment and electronics for electron microscopes. In 1956, he decided to move to the newly founded European Organization for Nuclear Research, i.e., CERN, in Geneva. At CERN, under the leadership of John Adams and Colin Ramm, Simon became involved in the technical design of the pole-face windings, multipole correction lenses, and their power supplies for the 28 GeV Proton Synchrotron, which was under construction at that time. Fifty years on, the PS is still in operation today, as the heart of CERN’s accelerator complex and feeding the SPS and LHC. Simon had a growing interest in the handling of particles. After working for a year in 1960 on a separated antiproton beam, Simon proposed the idea of a high-current, pulsed focusing device, the magnetic horn [2]; this was his first CERN invention, at the tender age of 37 (Fig. 1). Here, the charged particles traverse a metal wall a few millimetres thick, through which a pulsed high current flows. The original application of the magnetic horn was in the context of neutrino physics, where beams of pions have to be tightly focused. When the pions then decay into muons and neutrinos, an equally well-focused neutrino beam is obtained. The first horn [3, 4], for the PS neutrino beams in the early 1960s, was of monstrous size (Fig. 2), and today’s horns for the beams sent from the SPS to Gran Sasso, 732 km away, are equally huge. A much smaller magnetic horn was also essential to the collection of antiprotons from a production target and subsequent accumulation and storage in the Antiproton Accumulator (AA) in 1980; an improved 1993 version is still in use today at CERN’s Antiproton Decelerator facility. His interest in handling particles even extended to calculating the neutrino flux, taking the focusing provided by the horn into account [4, 5].

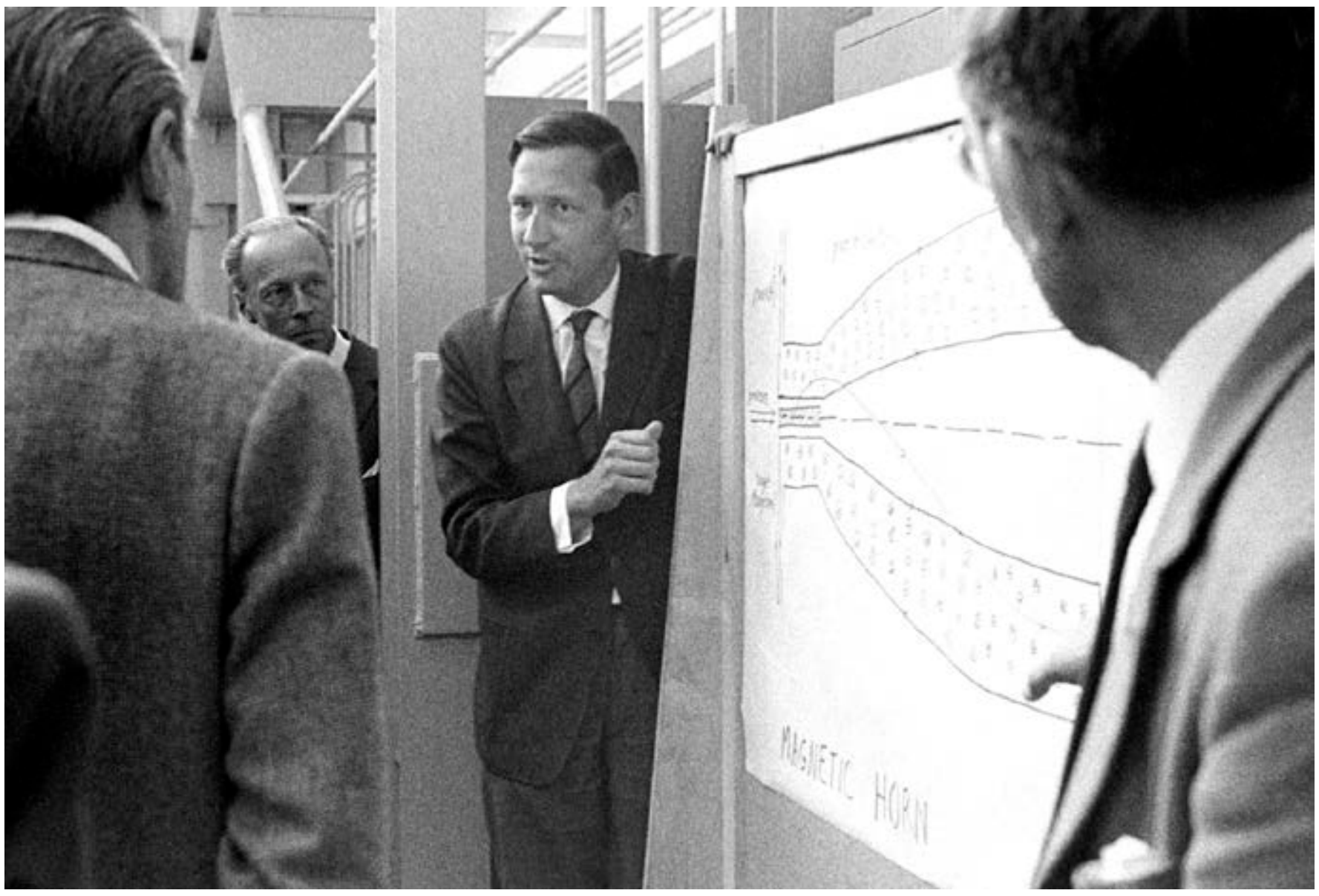

**Fig. 1:** Simon van der Meer, 37 years old, explaining the principles of a magnetic horn, June 1962

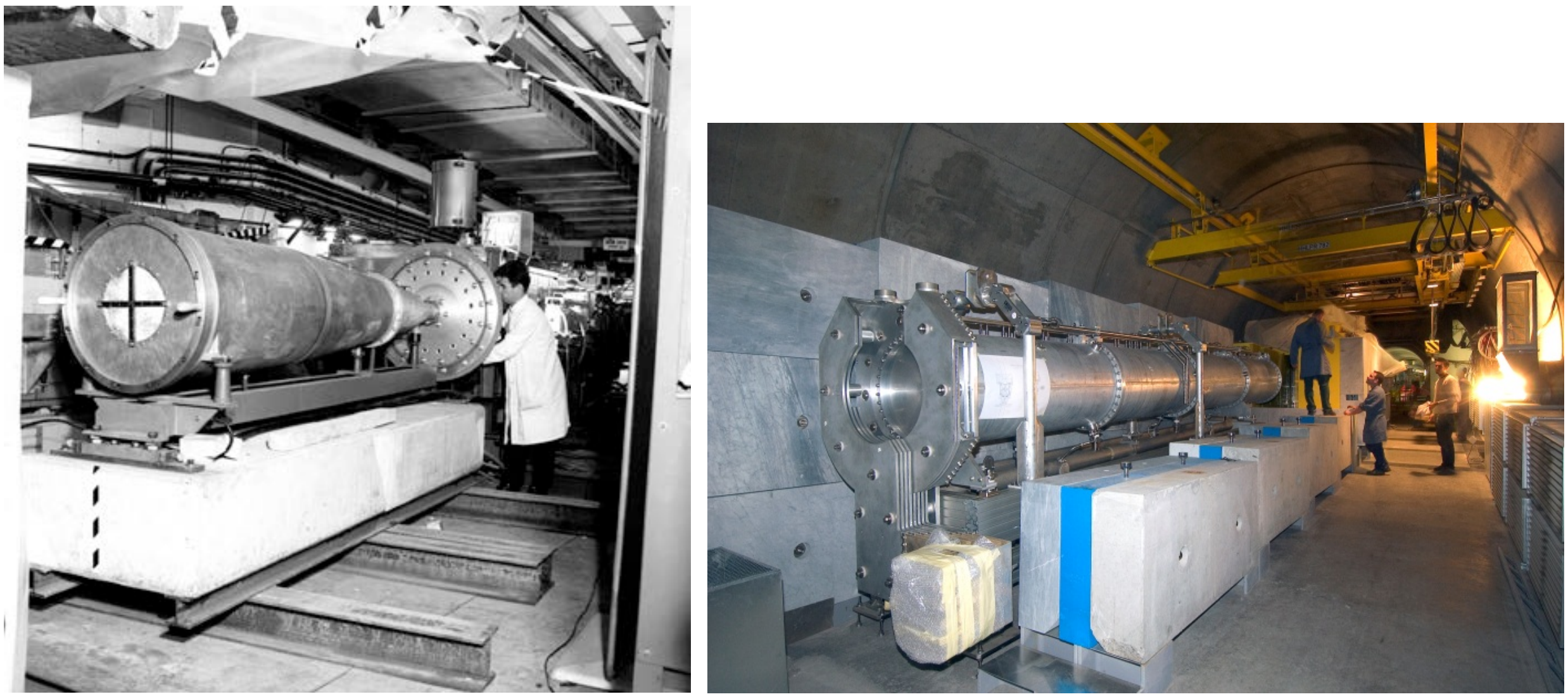

**Fig. 2:** The horn for the1963 PS neutrino beam experiment, and the 2005 version of the horn for the 450 GeV beam sent from the SPS to Gran Sasso

In 1965, Simon joined a small group led by Francis Farley preparing for a second experiment on the measurement of the anomalous magnetic moment of the muon. There, he designed a small storage ring (the ‘g-2’ ring) and participated in all phases of the experiment. As he stated later, this period provided invaluable experience not only in learning the principles of accelerator design but also in getting acquainted with the workings and lifestyle of experimental high-energy physicists.

It was during this period, in 1966, that Simon met his future wife, Catharina Koopman, during a skiing excursion in the Swiss mountains. Simon later described his decision to get married soon after as “one the best decisions of his life”, and they had two children, Esther and Mathijs, born in 1968 and 1970, respectively.

In 1967, Simon went back to do more technical work as opposed to particle handling or accelerator design, by becoming responsible for the magnet power supplies. These power supplies were for the ISR, under construction at the time. During the ISR period, he developed the two-dimensional magnetic-field calculation program MARE [6], which was used in magnet design for both the ISR and the SPS. Here too, Simon’s modesty meant that he did not want put his name first on the CERN Yellow Report, as recounted to me recently by Romeo Perin.

Soon after, the so-called 300 GeV project for the construction of the SPS at CERN was approved, under the leadership of John Adams; Simon was part of the SPS Design Committee, as head of the SPS power supplies group (Figs. 3 and 4). Here, he proposed that the generation of the reference voltages for the bending and quadrupole supplies should be based on measurements of the field along the cycle, and gave an outline of the correction algorithms [7]. His proposal resulted in the first ever computer-controlled closed-loop system for a geographically distributed system, as the 7 km circumference SPS was; this was a no simple feat for the early 1970s. Measurements of the main magnet currents were introduced only later, when the SPS had to run as a storage ring for the SPS $p$–$\bar{p}$ collider. Simon’s work on the SPS power supplies using digital-to-analogue converters, simple integrators, and a dedicated minicomputer remained the mainstay of the SPS all the way to 1996. He also left a large legacy of software in the SPS control room consoles for generating main-power-supply ramps, trimming the main dipoles, and adjusting the tune, together with archiving of power supply settings. Many of these pioneering tools and ideas were taken up or were integrated into other application programs for the SPS, and have now been inherited by the LHC.

Simon’s modesty is also reflected in another anecdote from that period, at the debut of the Lab II SPS 300 GeV project when the structure of the project was being set up, starting with the nomination of the Group Leaders. The story goes that when Simon was asked how many people he would need to set up the first team for the SPS powering scheme, including the 380 kV connection to the French electricity network and the machine power supplies, his answer was four or thereabouts. This was in striking contrast to a fellow compatriot who was asked to set up the beam transfer team for the SPS, and who apparently wanted 35 or more people!

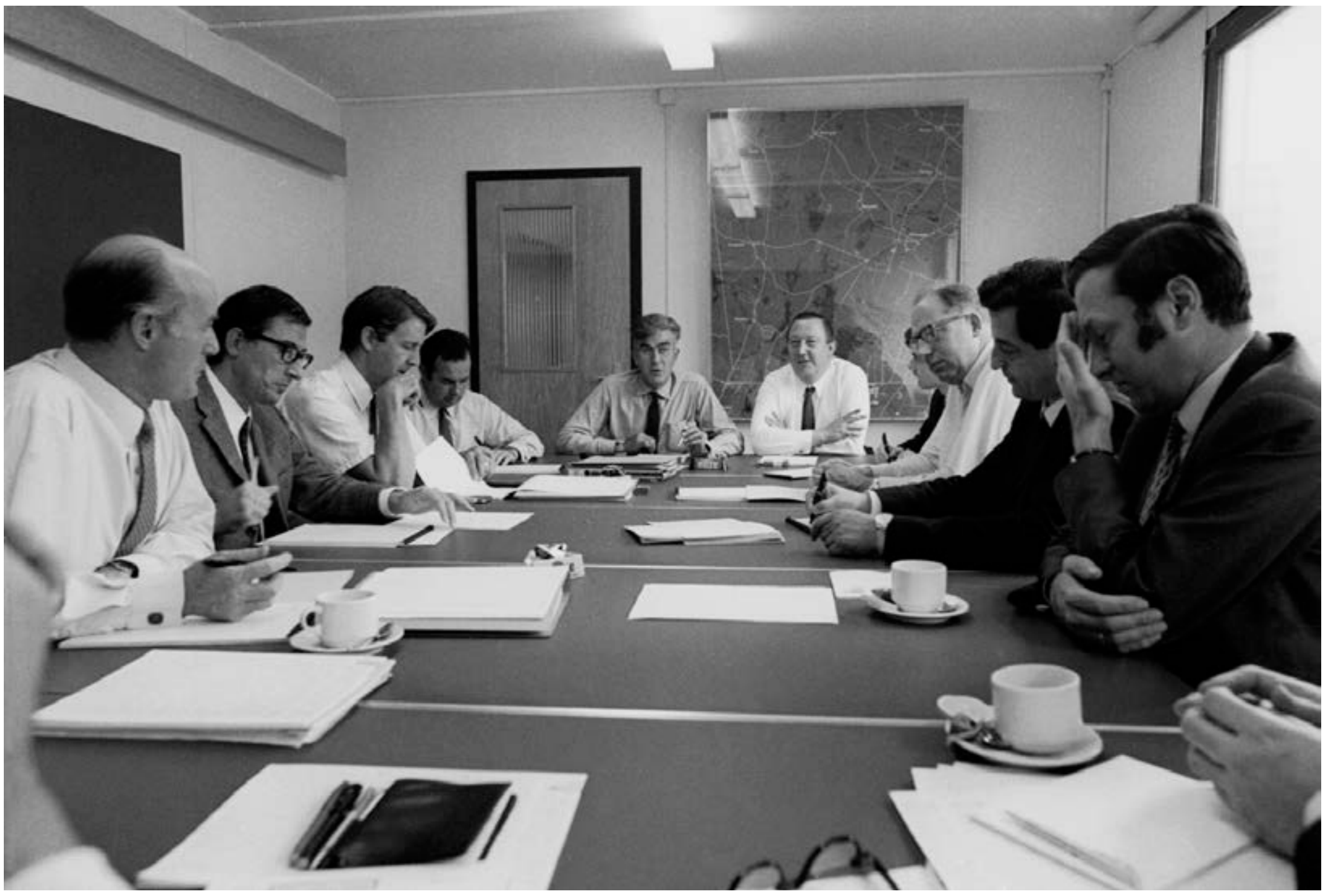

**Fig. 3:** The SPS 300 GeV Project Group Leaders in a Meeting; Simon van der Meer is first on the right. (October 1971)

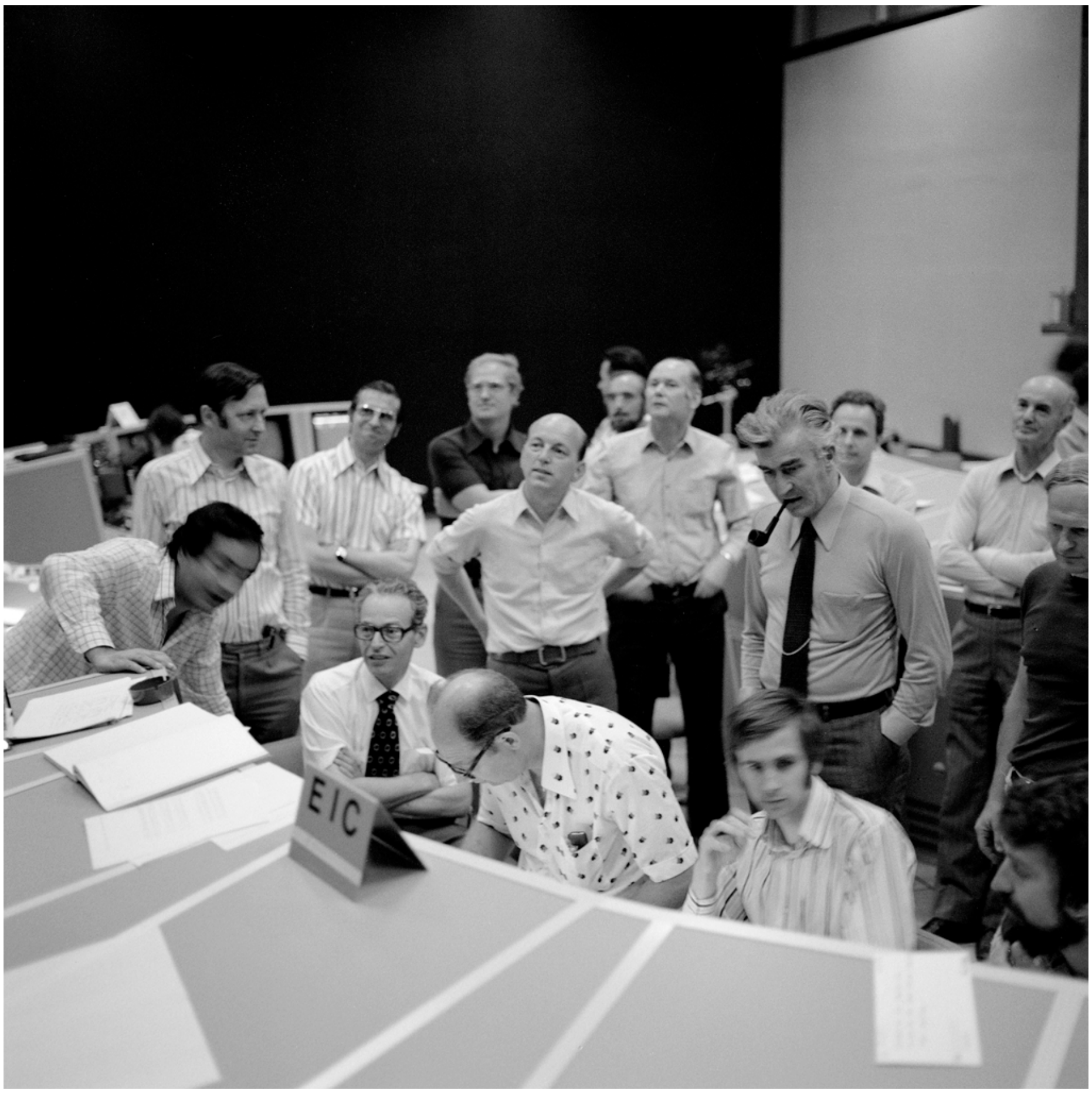

**Fig. 4:** The first beams at the SPS, summer 1976; Simon van der Meer standing on the left

It is also important to note that his work on the SPS power supplies [7] was written up by him in August 1972, the same month as the seminal note on “stochastic damping of betatron oscillations in the ISR” [8].

During his activities at the ISR, Simon developed a technique using steering magnets to vertically displace the two colliding beams with respect to each other; this permitted the evaluation of the effective beam height [9], leading to an evaluation of the beam luminosity at an intersection point (see Appendix A). The famous ‘van der Meer scans’ are indispensable even today in the LHC experiments; without these, the precision of the calibration of the luminosity at the intersection points in the Collider would be much lower.

It was in 1968, for the ISR, that a brilliant, new idea to increase the luminosity was conceived: the concept of stochastic cooling. “The cooling of a single particle circulating in a ring is particularly simple”, as Simon said in a qualitative description of betatron cooling for his Nobel Lecture in 1984 [10]. All that was needed was to measure the amount of deviation from the central orbit and correct it later with a kicker at a suitable location in the ring; the correction signal to the kicker had to arrive at a suitable location on the beam trajectory through the ring. However, the devil is in the detail for such a system. In reality, it is not possible to measure the position of just one particle, because there are many particles in the ring and there is electronic noise in the system, so that it is impossible to resolve a single particle. So, groups of particles, referred to as beam ‘samples’ or ‘slices’, must be considered instead.

For such a beam slice, it is possible to measure the average position during its passage through a pick-up and to correct for this when the same slice goes through a kicker. In addition, there is a need for ‘mixing’. Because there is a spread around the central momentum, some particles are faster and others are slower. This leads to an exchange of particles between adjacent beam slices, and this is vital for stochastic cooling. With a combination of many thousands of observations (from many thousands of turns), a sufficiently large bandwidth of the cooling system, low-noise or even cryogenically cooled electronics and good mixing, stochastic cooling *works*.

Simon only published the first internal note [8] on stochastic cooling in August 1972, apparently after much persuasion from W. Schnell, the leader of the ISR Instrumentation & RF group at that time. However, as Simon says in the final remark of this decisive note, he did this also because ‘fluctuations upon which the system is based were experimentally observed recently’, meaning early 1972; see, for example, [11] and, later, [12]. The first experimental record of the observation of stochastic damping (cooling) over ~4 h is in [13], from November 1973.

Figure 5 illustrates the first pick-up-to-kicker stochastic cooling system, used between ISR points 5 and 6, with a 120 m coaxial line. An experiment carried out using that system led to the results shown in Fig. 6, with a clear indication of a reduction in the r.m.s. amplitude. The effect on a stored beam of switching the stochastic cooling on and off, spread over 13 h, was more discernible in a test carried out in 1974 [14].

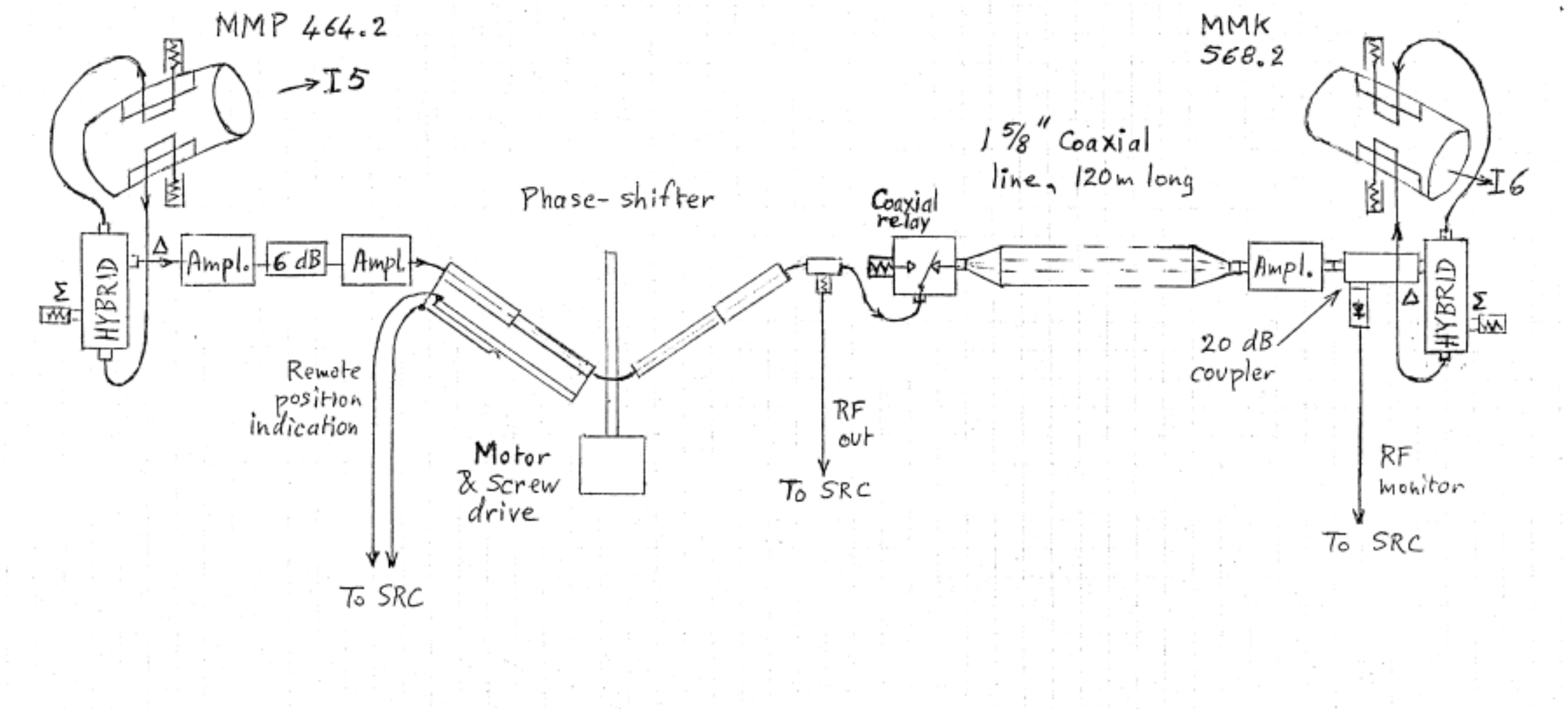

**Fig. 5:** The first stochastic cooling system used at the ISR (November 1973)

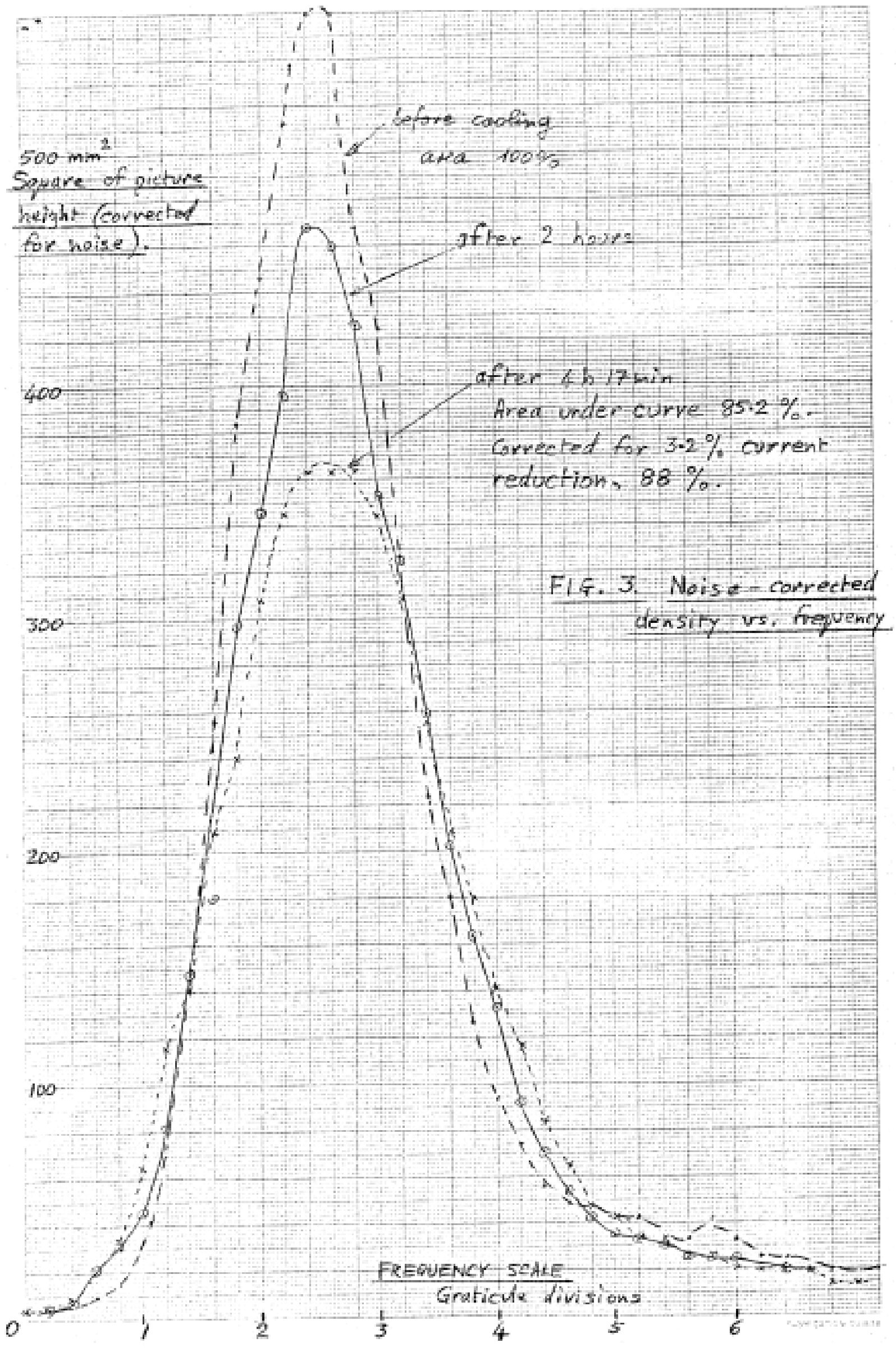

**Fig. 6:** Treated data from vertical Schottky scans after subtraction of the noise floor, 27 November 1973

## 3 The CERN proton–antiproton collider

In the late 1960s, a major challenge for particle physicists worldwide was the search for the W and Z particles—the carriers of the weak force—but no accelerator at either CERN or elsewhere provided enough energy to create these predicted particles.

In 1966, Budker at Novosibirsk suggested that high-luminosity proton–antiproton collisions in a single magnet ring might be made feasible by means of electron cooling. This was followed by the development and experimentation on stochastic cooling at the ISR in the early 1970s. This then led to the bold and imaginative idea of Carlo Rubbia in 1976 of using the CERN 400 GeV SPS or the Fermilab Main Ring as a single-magnet-ring p–$\bar{p}$ collider.

Head-on collisions between two beams provide the highest collision energies, a principle exploited at CERN in the 31 GeV ISR, in which two proton beams circulated in two interlaced rings. The new idea was that if a dense enough antiproton beam could be produced, then an accelerator with higher energy than the ISR, filled with a proton beam travelling in one direction and an *anti*proton beam in the opposite direction could be used as a collider.

To apply this idea to the SPS, antiprotons had to be provided in sufficiently high quantities, and this required a new, small storage ring, the AA. For CERN, this project was considered as an 'experiment' because of the multitude of challenges and unknowns that had to be overcome. The AA was indeed an adventure into uncharted territory. Never before had a project called for such imagination, involving the whole of CERN.

In 1976, two CERN working groups examined the technical aspects of such schemes and the physics potential. Finally, CERN decided to pursue two courses of action in parallel. One was to construct rapidly a small ring (the Initial Cooling Experiment, ICE) to study both electron and stochastic cooling; the other was to set up a study group to prepare a design for a p–p̄ facility using the SPS as a storage ring for collisions. Initially, the study group proposed using two separate rings for collecting and cooling antiprotons, because it was clear that the electron cooling scheme would only work at low energy for the large-emittance antiproton beam. Hence, the antiprotons were to be decelerated in the second ring. Meanwhile, the many experimental tests at the ISR, further theoretical developments, and, most importantly, the proposal of a faster and more efficient method of longitudinal cooling (the Thorndahl filter method, which uses a shorter, high-signal-to-noise 'sum' pick-up as opposed to a 'radial' pick-up) gave CERN the possibility of a solution based entirely on stochastic cooling and stacking. This, then, is how the AA was conceived and born—a fixed-field, single DC-operated accumulator ring. The potential saving in cost and complexity over the two-ring idea was the ultimate criterion, despite the fact that this solution represented three orders of magnitude of extrapolation from the ISR stochastic-cooling experiments.

(Just as a side remark, electron cooling works by transferring energy from an antiproton beam circulating in a storage ring to a 'cold' electron beam travelling in synchronism with the antiproton beam over part of its path. The electron beam has to be continuously refreshed from an electron gun. In other words, if precise matching of the velocities of the two beams is achieved, the Coulomb interaction tends to equalize their 'temperatures' so that an electron beam with very little transverse motion will reduce the transverse motion of the antiprotons, whose density is thus increased. However, electron cooling works best at very low energy, whereas for the stochastic technique, the cooling rate is independent of energy. As mentioned earlier, it was also the fast and efficient pre-cooling of the momentum spread obtained by Lars Thorndahl's filter method that clinched the decision to use exclusively stochastic cooling for the p–p̄ collider project. Electron cooling also needs a rather large amount of well-functioning, reliable equipment.)

The AA proposal called for an overall increase in the antiproton density from the production target to the stack core by a factor of over $10^9$. Meanwhile, in 1978 the ICE ring gave encouraging results for stochastic cooling [15], confirming cooling in all planes, though at timescales (longitudinally) of the order of 10 s, still about 30 times slower than what was required in the AA. However, the stochastic stacking process, which was prone to instabilities (as first pointed out by Sacherer [16]), was an essential feature of the accumulation scheme; this could not be tested in the ICE. The process involved simultaneous cooling in the two transverse planes and increasing the longitudinal density by four orders of magnitude while the particles were moved into the dense core, using a combination of filter-based and radial-pickup-based Palmer cooling techniques to avoid instabilities. This, then, was the biggest gamble in the launch of the AA, because it could only be studied in detail by theoretical calculation. Fortunately, the AA performed as expected in this respect.

I first met Simon in the heady days of mid-1980 when the AA had been just been constructed in the record time of two years after approval, under the inspirational leadership of Roy Billinge and with Simon as a joint Project Leader. This was the era when the AA project team had meetings every Friday morning, where no hierarchies were discernible (a far cry from today's CERN) and everyone from the most junior technicians to the senior accelerator physicists got together and discussed the weekly issues. If Simon spoke in dissent or agreement, his views would carry the day, and that was how things would be implemented; as a newcomer to this august crowd of accelerator builders and experts, I found it quite an eye-opener to be part of this very vibrant and stimulating group of CERN staff of many nationalities, working across many Divisions. The meeting would then adjourn to Tortella's, CERN's No. 2 restaurant, where Roy Billinge had an astute arrangement, with food and drinks flowing for a fixed modest sum. Simon was a regular attendee, participating actively in over-the-table discussions of all matters of work and otherwise.

These early AA days were also a period of prolonged controversy about control philosophy, such as the choice between simple, SPS-style touch terminals directly connected to the AA control computer and a two-layer model with control room consoles-computer connected to the front-end accelerator-related computer via a proprietary network. Simon's insistence on direct AA controls and facilitated interpreter (NODAL) programming carried the day. This was also the era when Simon gradually took over all the AA application programs, written initially by equipment builders and others; the stand-alone HP computers and other independent beam measurement instruments were frowned upon, and Simon succeeded in eliminating all such devices by insisting on direct connection to the AA computer via the established CAMAC interfaces.

Issues of operation centred on the choice of whether to use the local AA control room (Figs. 7 and 8) or the central Meyrin control room, and how to staff the delicate operation of the AA, with accelerator experts or Operation Group staff. After all, the AA was built as an 'experiment' with the help of people from seven different CERN Divisions! Simon's prolific programming ability meant that by 1981 he had implemented virtually all of the user-friendly tools required to allow the operation of the AA to be performed by the Operations Group team that had been put together by then.

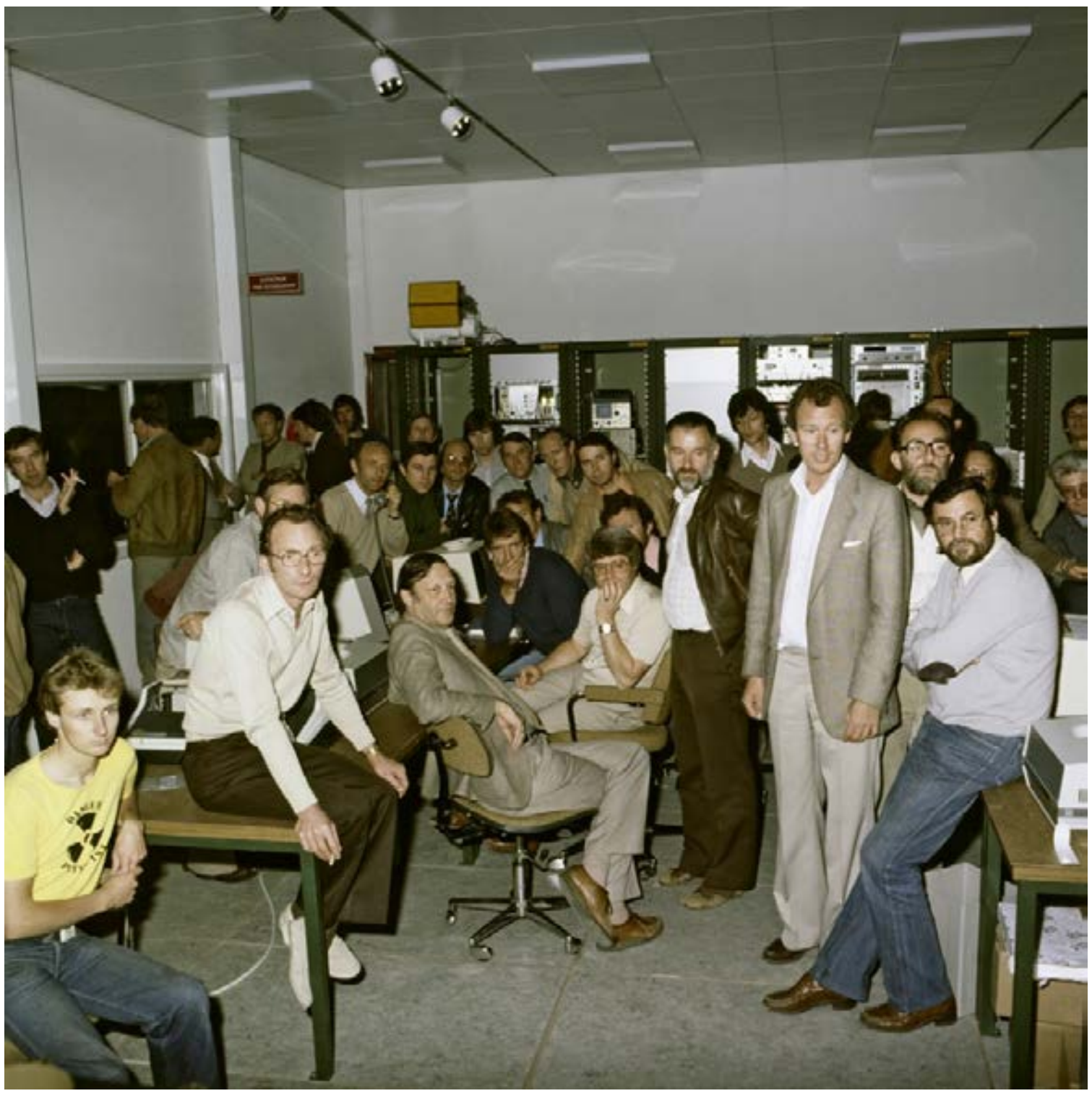

**Fig. 7:** First proton test beams for the AA, July 1980, with Simon seated on the chair middle of the photograph, waiting to see the beam flash on a scintillation screen!

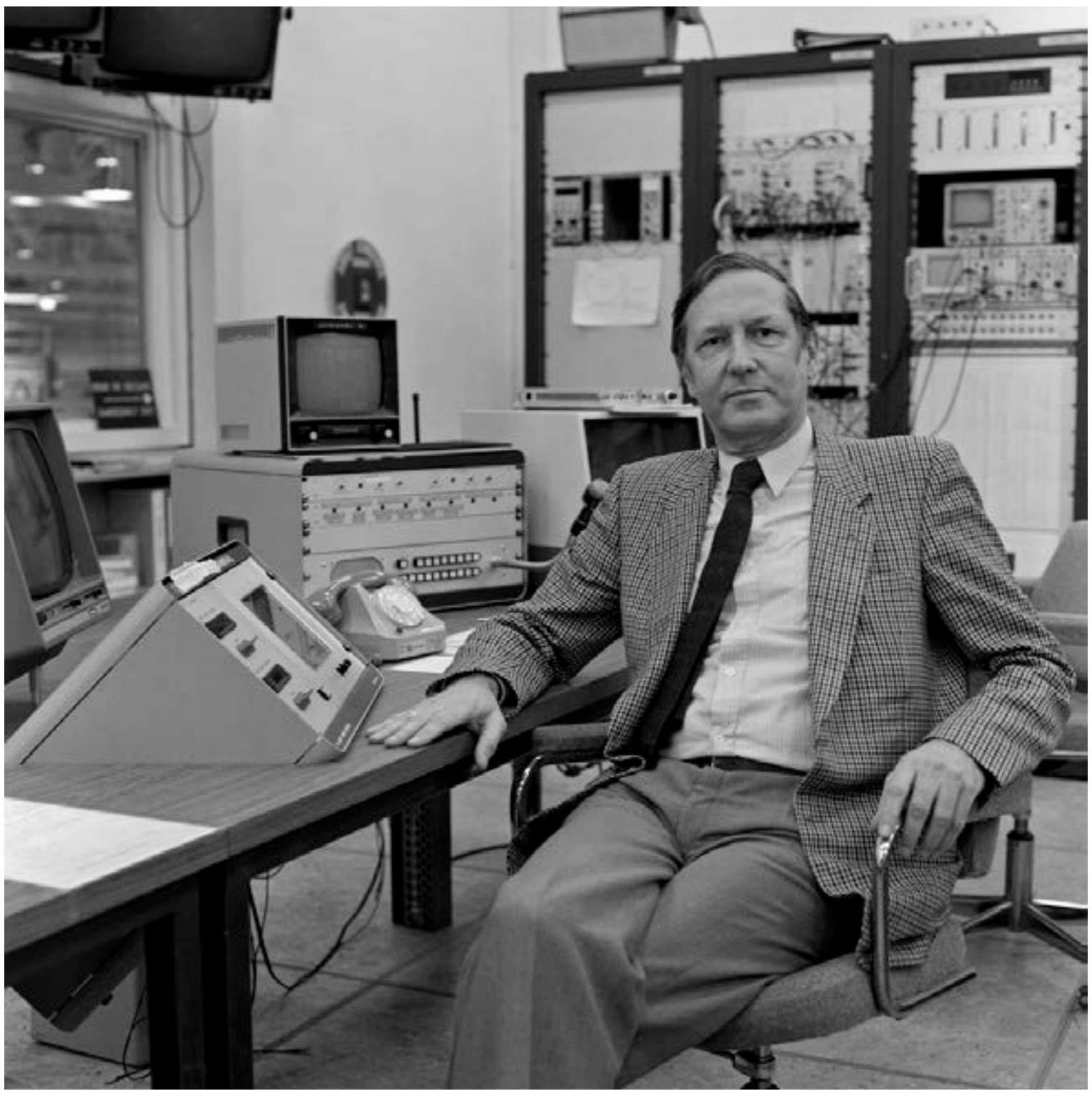

**Fig. 8:** Simon van der Meer in the local AA control room with the SPS-style touch terminal console, January 1984

However, for the relatively slow stacking and accumulation process, with a night shift at the AA staffed by a single operator, this implied putting an SPS-style touch-screen console in the Meyrin control room (MCR) so that the AA night shift operator could be with his colleagues in the MCR. This move of an SPS-style console into the MCR was considered sacrilege by the host Division's controls group, not to mention the Operations Group; their mantra was 'giant consoles banalisé' (i.e., all-purpose consoles) and an equipment reservation system, which was anathema to Simon and his ways of working from his SPS days! The reservation system implied tripling the programming required for controlling a single piece of equipment (reserve, use, release); one has to realize that the SPS had been commissioned just a few years earlier (in 1976) under the formal title of another laboratory (CERN Lab II), on the other side of the border, with all the CERN cultural and interdivisional rivalries of the era. Simon's applications, he insisted, would run directly on the AA computer only, using the SPS touch screen, without the reservation system and without network overheads, this being in the pre-Ethernet era. His insistence swayed the decision to permit the delicate operation of the AA to be done from the MCR using the SPS-style touch screen, thereby relieving the members of the project construction team of routine operation duties. This dedicated touch terminal remained the bastion (and bane to the PS Controls Group) of p–p̄ operation until early 1991. An emulator then took over, running on the MCR X-Terminal consoles that were then the standard provided by the Controls Group, long after the touch terminal had first provoked the wrath of the Operations Group. It is interesting to note that the leader of the Controls Group at that time was a contemporary of Simon's, and was a compatriot as well as a close friend privately.

For the aficionados of today's touch-screen smartphones & tablets, it is of interest to note that we at CERN were the pioneers of this method of human–machine interaction in the early 1970s. When pull-down menus and mice came into common use in our control rooms around the 1990s at the expense of touch screens, it was considered a step backwards by many!

There were also proxy wars about local AA control room devices and instruments, whether they were essential in the local control room or not. The final, de facto decision was Simon's: he decided whether a device would be integrated into his sophisticated application programs or not. In the latter case, the use of that instrument would fall into abeyance, much to the chagrin of the initial provider. My direct work dealings with Simon started during this period. I had to juggle the wishes of Simon versus the PS Division's Controls and Operations groups or the providers of dedicated beam measurement instruments. When we came to the first delivery of antiprotons (to the ISR in early 1981, and to the SPS later), we had an initial concern, due to Simon's insistence on decoupling the AA from the rest of the PS or SPS park of Norsk-Data computers. Hence, intricate handshake procedures had to be formulated [17] and put in place.

Simon was indeed a hands-on man. He would often agree to work night shifts to perform machine improvements or experiments, particularly if he deemed it necessary to work quietly, without the usual disturbances of day shifts and with more steady availability of test beams from the PS. A case in point was certain corrections on magnets where he was the unique expert. The AA and, later, the AC (Antiproton Collector) both had many packs of washers bolted onto the quadrupole end shims for calibrating the gradient (Fig. 9). Based on beam measurements, Simon would meticulously calculate a new configuration of washers. Simon and I worked on such night shifts on several occasions, where I would assist Simon with the tedious and time-consuming process of removing or adding washers from or to all the relevant magnets and performing the measurements.

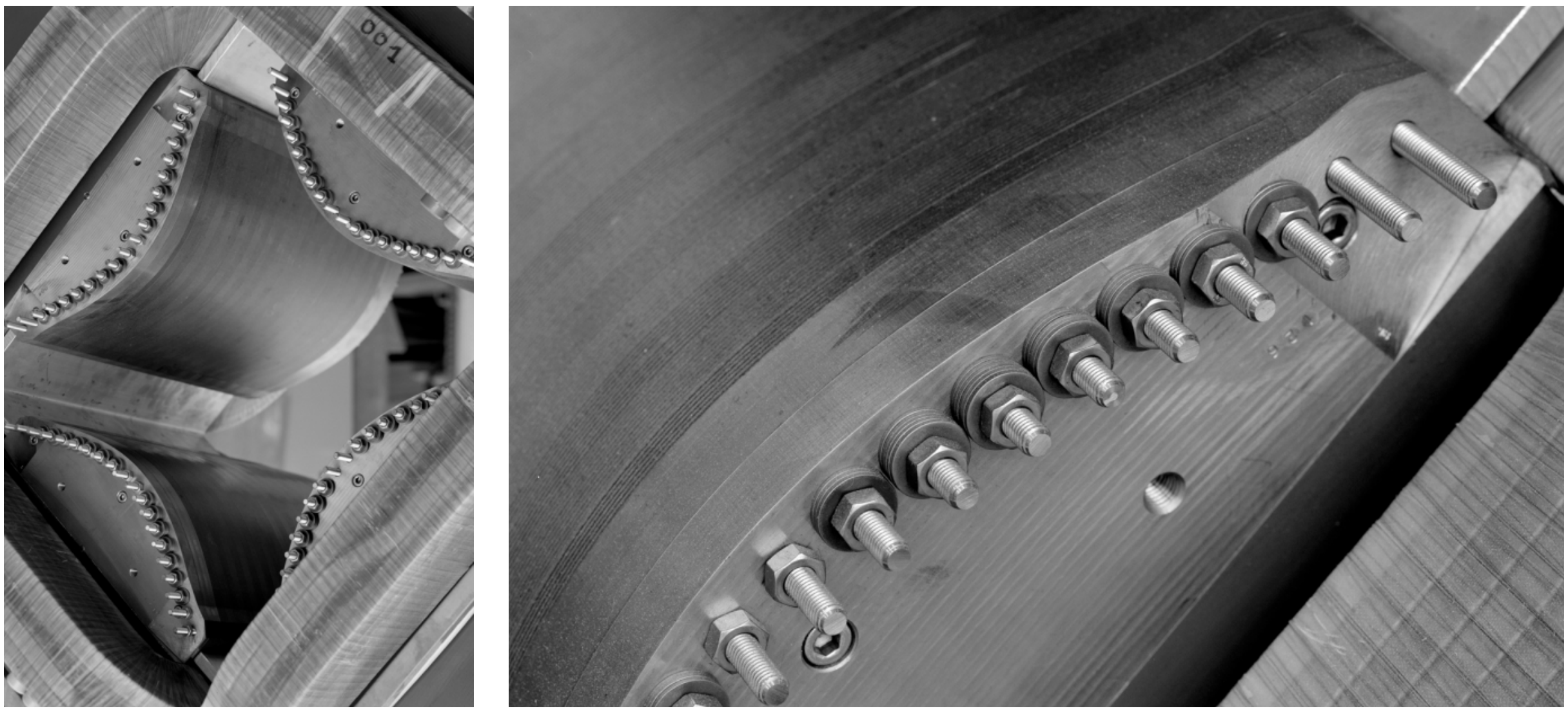

**Fig. 9:** Narrow quadrupole at the AA with shimming washers, and a pole end in detail

We would make the beam tune measurements, recalculate the washer configuration, make the machine electrically safe by manually locking out the magnet circuit breakers, enter the machine, and modify the washers; then it was back to the control room via the wash room to clean off all the graphite grease from our hands, unlock the circuit breakers, switch on the magnets, and request a proton test beam from the PS. The injected beam would then be captured by the RF system at injection and moved across the aperture to the desired momentum position and the tune re-measured, using transverse Schottky scans. Simon would then recalculate another washer configuration, and the whole process would start all over again. Simon's hands-on versatility meant that he even showed me how to get the black grease off my hands most efficiently, without recourse to water!

The goal was, of course, to obtain the desired tune across the full momentum bite of the ring. Finally, Simon would fastidiously note all the results in the paper logbook in the control room.

Whereas others calculated and noted the AA or AC parameters to five or more decimal places, Simon knew where three would do and had it all in his head.

Simon had a vested interest in the AA machine being perfectly tuned, and he was not going to leave that job to anyone else. He applied this philosophy with equal vigour to the AC ring, which we commissioned in 1987. He would work at the highest computational level and then apply himself to the manual tasks with equal dedication. He never complained; it was a job that had to be done right, and the only way to achieve that was to do it himself. So that is exactly what he did. Of course, he fully expected everyone around him to be equally committed. Not always an easy act to follow!

After Roy Billinge became the PS Division Head, he usurped the AA Group's Friday morning meeting slot and lunch, and we had to move to Mondays for weekly meetings and lunch, which too moved to the local pizza restaurant across the road from CERN. The lunch gatherings had a much reduced number of participants, but included the ebullient and charismatic group leader, Eiffionydd Jones. The success of the AA meant that a series of Fermilab scientists took part in these meetings and lunches. Particular names that come in mind are Rol Johnson, Gerry Dugan, John Marriner, Carlos Hojvat, and occasionally John Peoples, Jack McCarthy, Alvin Tollestrup, and Fred Mills.

Fermilab too went through the idea of using moving shutters inside the accumulator ring, similar to those in the CERN AA, which moved in every 2.4 s to decouple the cooling systems, i.e., to protect the accumulated stack from the fast, powerful pre-cooling systems. Electron cooling was also discussed at Fermilab; eventually, however, the Fermilab $\bar{p}$ source took the final shape of a debuncher ring and an accumulator ring, with higher-frequency-band stochastic cooling systems only. Many people from Fermilab sought Simon's counsel, as did others at CERN in those days. The then young physicist Jeff Hangst (recently of antihydrogen fame at the CERN AD facility) would not have forgotten his first encounter with Simon on visiting CERN, which happened while he was reporting on Fermilab's ideas of using a 'SEM grid' before the production target receiving the 120 GeV beam. As Jeff would recount later, Simon was not shy to castigate him, albeit discreetly, calling the idea 'fiction' because of the resolution of the grid wires, much to Jeff's astonishment! The technical notes and reports of that era for the Fermilab $\bar{p}$ source reflect very much the notes and reports for the CERN AA, sometimes even down to using the notation $q$ instead of $\vartheta$ for tunes; see [18] for an example.

This period was the beginning of the very fruitful CERN–Fermilab collaboration in the antiproton accelerator community. Ideas and people crossed the Atlantic, and eventually ideas even came back, such as CERN's construction of the larger-acceptance AC ring for fast 'pre-cooling' systems, the abandonment of shutters in the AA, and the advent of higher-frequency-band stochastic cooling systems at CERN. I believe Simon did not like to travel very much. On one occasion, I remember that he agreed to travel for a review or something similar and the meeting was held at Chicago O'Hare Airport, from where he took a flight straight back. This flight had to be via Amsterdam so that he could buy books in his mother tongue at the airport!

## 4 Nobel Prize

The first p–$\bar{p}$ collisions at the SPS occurred in July 1981 and the first real period of physics runs took place in 1982. December 1982 saw the Collider reaching an integrated luminosity of 28 inverse nanobarns and Carlo Rubbia offering a champagne-only party with 28 champagne bottles! It suffices to say that the first signs of the W boson were announced soon after, in January 1983. This was to be followed by the discovery of the Z (Fig. 10), announced in May 1983. I shall not go into too many details of the discoveries of the W and Z and the Nobel Prize because these have been amply recorded elsewhere, both at the time and later.

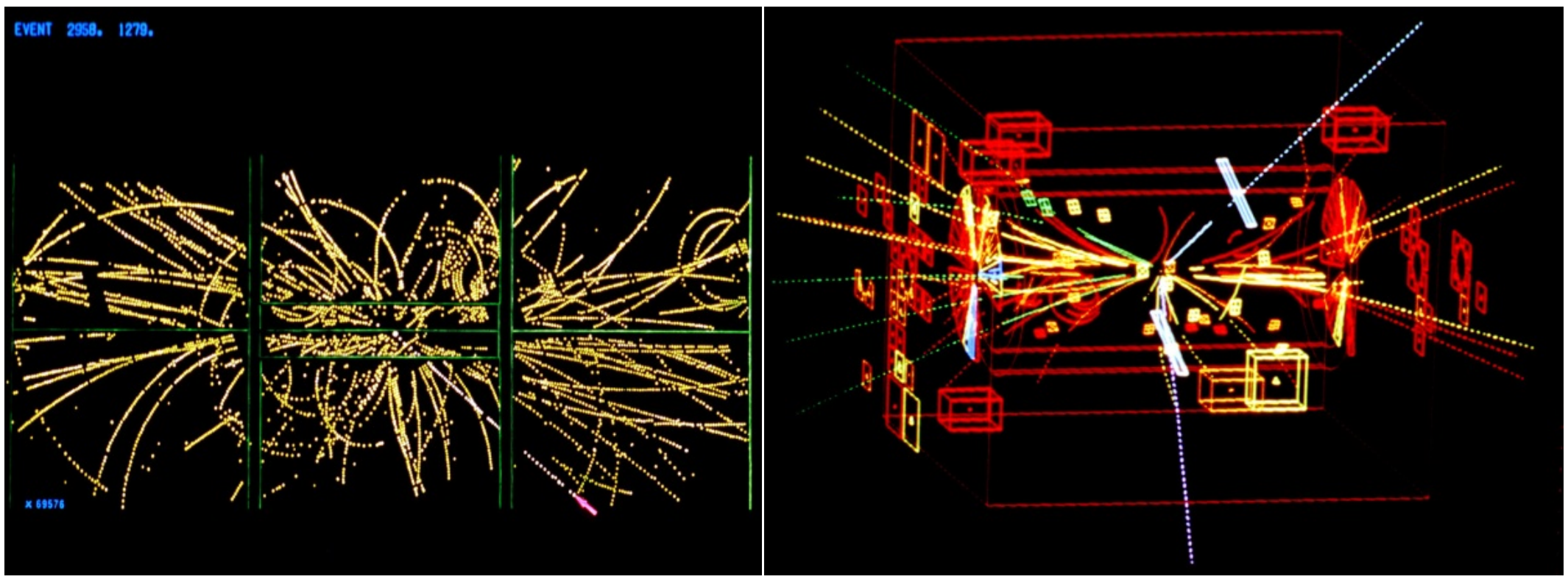

**Fig. 10**: First W event, December 1982 (left), and first Z event in the UA1 experiment, April 1983

When the telex (Fig. 11) came on 17 October 1984, Simon was sitting in his office in Building 19 and, characteristically, Carlo Rubbia was in a Milan cab on the way to Milano Linate airport en route to Trieste for a conference. The story goes that the taxi radio was interrupted by a news flash because of an Italian sharing the Nobel Prize in Physics. Initially, Rubbia's excitement was hardly believed by the taxi driver but, after persistent efforts to convince him, Rubbia got a free ride!

The Nobel citation read that:

> The Nobel Prize in Physics for 1984 was awarded jointly to Simon van der Meer and Carlo Rubbia for their decisive contribution to the Large Project, which led to the discovery of the field particles W and Z, communicators of weak interactions.

There was euphoria in the control rooms, particularly the local AA control room, where we gloated over the telex copy and quickly stuck it in the AA logbook, and champagne started flowing. There were similar scenes in the UA1 cavern and control room (Fig. 12). The antiproton accelerator community certainly liked the words 'large project' in the citation, because it gave a collective recognition too! It makes one wonder what phrase the Nobel committee will need to invent if prizewinning discoveries are the LHC's forte.

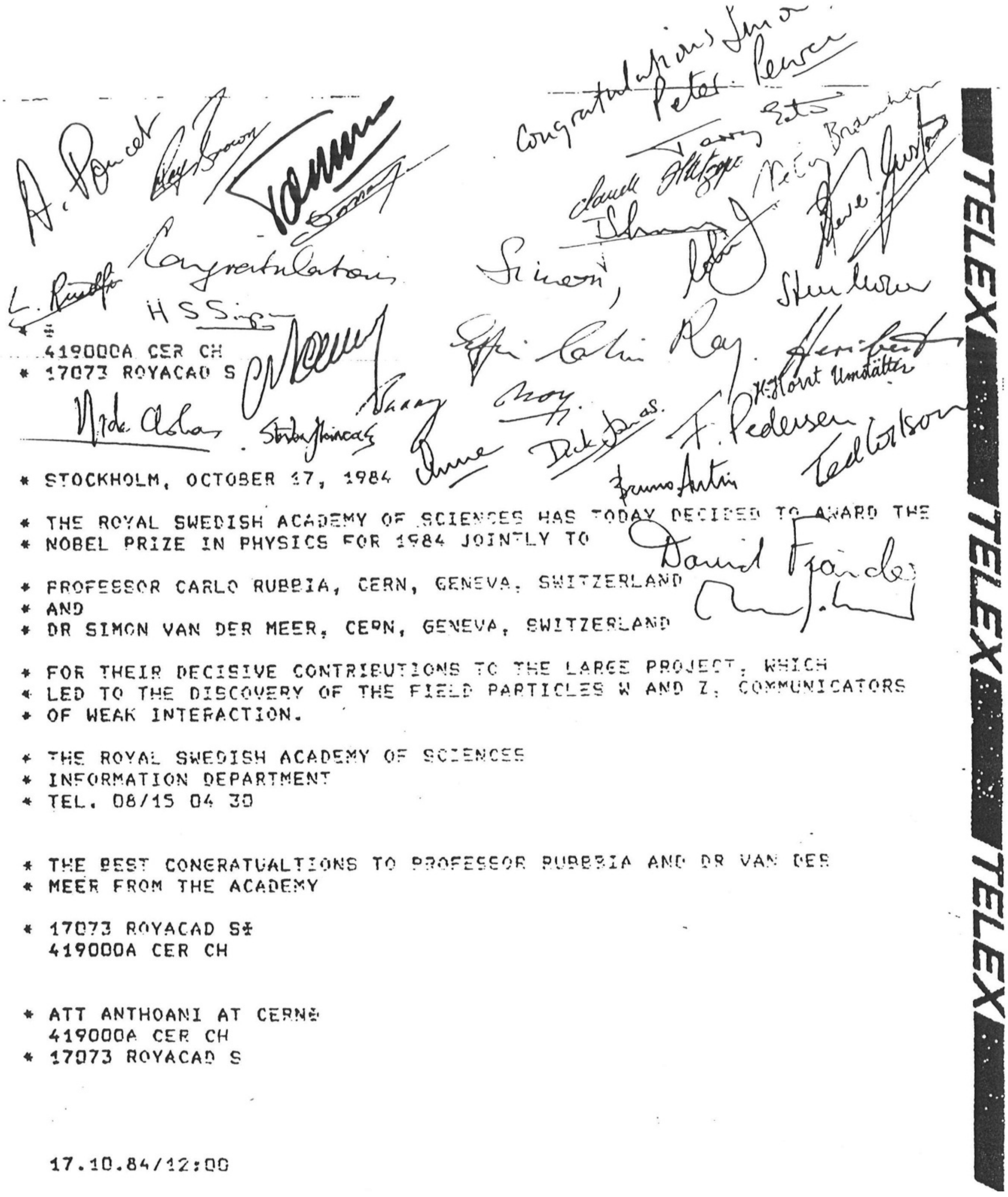

419000A CER CH
* 17073 ROYACAD S

* STOCKHOLM, OCTOBER 17, 1984

* THE ROYAL SWEDISH ACADEMY OF SCIENCES HAS TODAY DECIDED TO AWARD THE
* NOBEL PRIZE IN PHYSICS FOR 1984 JOINTLY TO

* PROFESSOR CARLO RUBBIA, CERN, GENEVA, SWITZERLAND
* AND
* DR SIMON VAN DER MEER, CERN, GENEVA, SWITZERLAND

* FOR THEIR DECISIVE CONTRIBUTIONS TO THE LARGE PROJECT, WHICH
* LED TO THE DISCOVERY OF THE FIELD PARTICLES W AND Z, COMMUNICATORS
* OF WEAK INTERACTION.

* THE ROYAL SWEDISH ACADEMY OF SCIENCES
* INFORMATION DEPARTMENT
* TEL. 08/15 04 30

* THE BEST CONGRATUALTIONS TO PROFESSOR RUBBIA AND DR VAN DER
* MEER FROM THE ACADEMY

* 17073 ROYACAD S+
419000A CER CH

* ATT ANTHOANI AT CERN+
419000A CER CH
* 17073 ROYACAD S

17.10.84/12:00

**Fig. 11:** Copy of the Nobel telex in the AA logbook for 17 October 1984

The success of the CERN antiproton adventure meant that the AA team received some consolidation money, managed parsimoniously by Ted Wilson. 100 MHz sampling transient recorders with sufficient memory, based on CAMAC, were just about available then from LeCroy. It was Colin Johnson's idea to invest in these. Dollarwise, they were very expensive, and even some detector physicists on the UA2 experiment were envious of our purchases and borrowed them for trials. Simon had been highly irritated with another commercial device, much more pricy, that was connected via a GPIB interface to the control system. I suspect it was purchased without Simon's tacit benediction! Hence began my next strong working collaboration with Simon. We worked for months together over this cause until we got the digitizer system fully operational according to Simon's wishes, and to get the maximum from the faster middleware software, where I had to negotiate with the usual groups concerned. Subsequently, Simon would invent more and more applications using digitized signals from position pick-ups and the like, until he retired. The other GPIB device went into its natural obsolescence soon afterwards! Not much later, such digitizers provided the backbone for LeCroy's pioneering digital oscilloscopes.

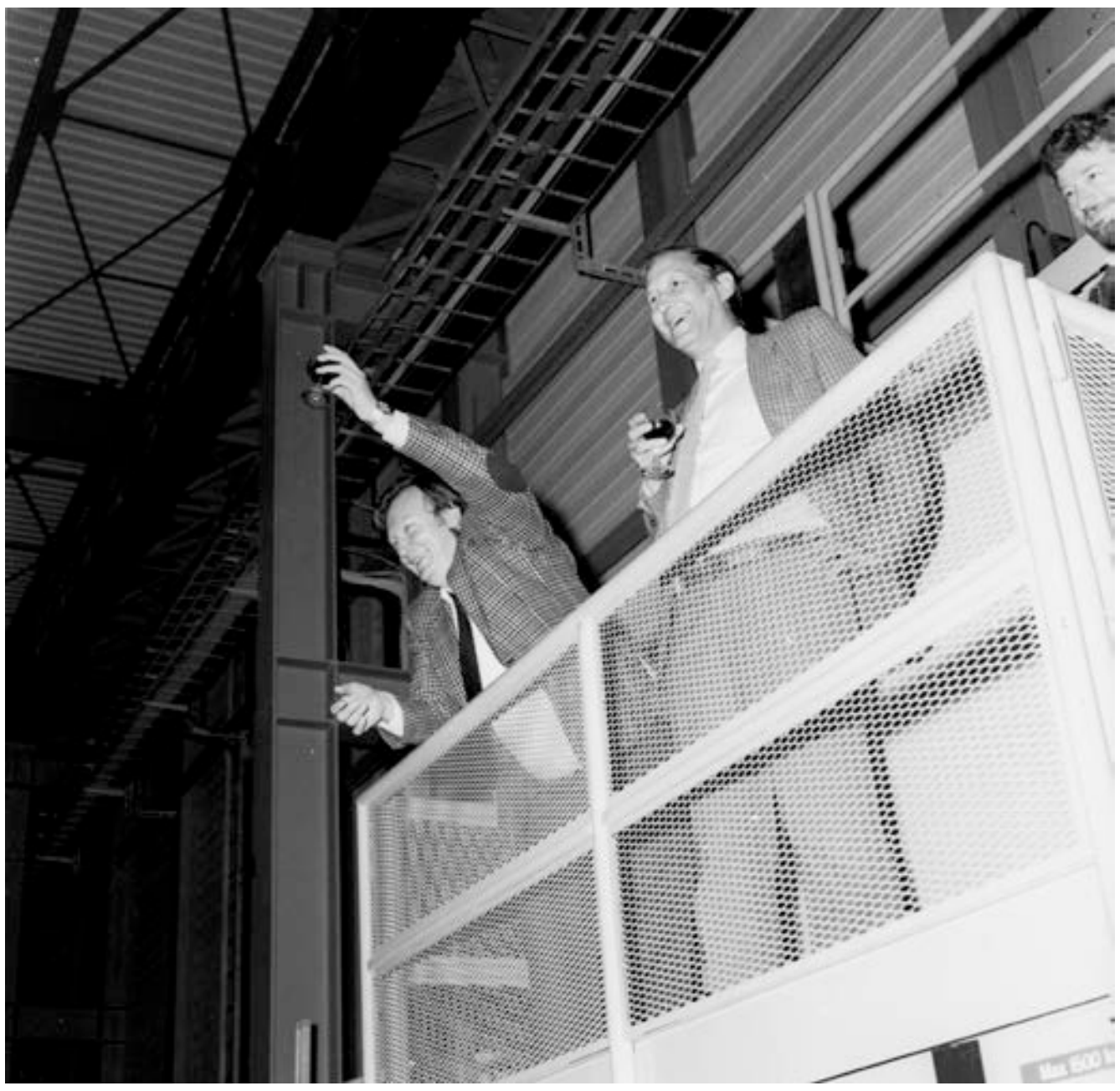

**Fig. 12:** Carlo Rubbia and Simon van der Meer cheering in the UA1 cavern, celebrating the Nobel Prize, October 1984

Perhaps a bit differently from today's LHC and today's huge experimental teams, there was much more cohesion between the UA1 and UA2 experimental physicists and the accelerator community. We used to have daily animated five o'clock meetings, but Simon refrained from attending them. Even though he was very much involved in daily operations, he preferred others to take up the AA cause for Collider operation.

Nobel Prize apart, the AA was not without its gremlins, which hounded us to the bitter end. The machine acceptance doggedly hovered around the 80π mm·mrad mark in both planes instead of the 100π foreseen, and the $\bar{p}$ yield achieved remained frustratingly low compared with design. This led to innumerable studies and proposals for yield improvement using ideas for better target assemblies, conducting targets, Li lenses for collection, and eventually the design and construction of the larger-acceptance AC ring in 1986–87, surrounding the AA ring.

The ACOL project, as it was called, required all the usual services and equipment, particularly the new beam instrument devices and their integration into existing control systems. Of course, it goes without saying that Simon would insist on everything following the same standards as the AA, so that we could adapt or augment the necessary hardware and software easily. The Project Leader was Eiffionydd Jones, who had a way of dealing with everyone, including Simon; he gave an explicit instruction that I should keep Simon happy and ensure that all went well during the planning, construction, and implementation phases for all of the equipment integration, including the beam instrumentation and control issues. We also had only 11 months for installation activities in 1986–87, with start-up and commissioning in June 1987. We were late with full machine equipment installation by June 1987 but kept to that date; the outstanding installation work was done during the daytime and commissioning at night. Simon and I were the two people fully excused from night shifts so that we could iron out all of the problems that had been exposed in the commissioning night shift with test beams. Simon and I had a tacit arrangement to go home in the evening as usual but come in again later, to see the fruits of our daytime modifications with evening beams until late into the evening. The AA and AC $\bar{p}$ source complex remained from 1987 to 1996 the most highly automated set of machines [19] in CERN's repertoire of accelerators, again thanks to Simon.

The success of the p–$\bar{p}$ collider programme also had a spin-off in the form of antiproton physics at very low energies. The Low Energy Antiproton Ring (LEAR) came into being at CERN in about 1983, using both stochastic and electron cooling systems. Antiprotons were extracted and sent to experiments in the CERN South Hall, Building 150. Simon's invention of noise-assisted slow extraction (so-called stochastic extraction) permitted spills for antiproton physics for up to 24 h, a boon for low-quantity antiproton trapping. Prior to that, low-ripple spills could last a few seconds at best. The CERN LEAR programme ended in 1996, but not before culminating in the creation of the first antihydrogen atoms in 1995 and the subsequent conversion of the AC ring into the Antiproton Decelerator, a new facility to continue the low-energy antiproton physics experiments.

Stochastic cooling systems became operational at Fermilab in the early 1980s and at GSI Darmstadt and Forschungszentrum Jülich (FZJ) in the early 1990s. Today, there are some nine or so other rings using stochastic cooling, all over the world.

## 5 Retirement

A few months before Simon was due to retire in December 1990, the Division Head, Roy Billinge, organized a meeting with Simon and me to cater for Simon's software legacy for the future operation of the $\bar{p}$ complex. In the following months, Simon meticulously documented the programs in which he had direct involvement and about which he had direct knowledge. These were early days for word processors and PCs, and everything was done very clearly as text files in his favourite interpreter. This, then, was the legacy he left to me, up to the last days of the AA (and AC) in 1996, when the $\bar{p}$ programme in that form ended at CERN. On his last day at CERN, he took the large output listings of his entire programs home, and I was welcome to call him if I needed him to look up anything.

He cleared out his office and insisted on not keeping a foot in CERN, unlike many others, then and subsequently. That showed the modesty and spirit that he had expressed throughout his career at CERN. He had done his bit; now it was time to do other things, including his voracious reading!

All those years, our Monday lunches had continued until the regrettable and early passing away of Eiffionydd Jones in early 1990. By then, the lunch participants were down to four or five; others had dispersed or disappeared to other activities. At the Monday lunches, Simon was always well informed about current affairs, including, surprisingly news from Britain; I found out later that this was because he bought the English Sunday paper *The Observer*. I once chided him, saying that the *Sunday Times*, which I read on Sundays, was of much superior quality! Little did I know that the crossword puzzle in the *Observer* was his raison d'être; he would usually succeed in completing it in no time on Sunday afternoon, explaining his excellent command of English.

After Simon's retirement, it was the AA group's veterans and Antipodeans Colin Taylor (an Australian who built the first CERN linac as well as the AA core cooling systems) and Ray Sherwood (a New Zealander of linacs and beam lines fame), Simon, and I that held the fort for pizza lunches in local establishments. Later, Flemming Pedersen, on his return from SLAC, joined us regularly, and Ted Wilson would join in occasionally, in between his CERN Accelerator School jaunts. There were others too, such as Horst Umstaetter, who became active semi-regular participants. Satellite dishes were in vogue in those days, and both English and Dutch channels were just about receivable in Geneva using different satellites and orientations. Simon's first retirement project was to build an orientation control for an 80 cm dish that he could operate from his living room. I remember the related over-the-pizza discussion, and within few days, the whole thing was functional. The dish was in the garden, and he had made a flat surface with a cable trench neatly covered to avoid mowing problems. I found out later that he had built a motorized antenna holder system in his usual efficient but frugal style, using a wheel taken from an unwanted garbage container, a barbecue grill motor, and a Tupperware box to keep out humidity!

In the early part of his retirement, Simon agreed to assist in some worthy cause, if my memory is correct. This involved writing and posting letters for a worldly cause and something to do with the Rio 1992 UN conference on sustainable development, the famous Rio92 climate change conference. However, it would be more appropriate to say that unlike many other Nobel laureates with more exuberant personalities, he preferred not to proffer counsel or take up causes, of whatever noble or laudable aim, because it would just be against the grain for him to do so. He was not a man in quest of self-serving platitudes.

Late in 2003, in preparation for CERN's 50th birthday in 2004, it was decided to compile and publish a 50th anniversary glossy book, with 50 chapters. Each chapter denoted a year and notable events in that year, in the history and life of CERN. I was asked to approach Simon to ask if he would write the 1980 chapter, on the beginnings of the antiproton adventure, just as Carlo Rubbia had agreed to write the 1983 chapter, among other personalities of CERN spanning fifty years. It goes without saying that Simon gave a characteristic response: negative, because it was all a long time ago, he did not remember much, and so forth, his usual modesty-driven reasons. After iteration or two, he agreed that if I wrote it, he would read and correct it. This, then, is what was published in [20].

In May 2004, I suggested to Simon that he should come and visit the LHC Magnet Test Facility, where I was responsible for the 24/7 operation to test all of the LHC cryomagnets. This he did after one of our lunches, and it was one of his last visits to CERN (Figs. 13 and 14). In the following years, back problems meant that he was reluctant to come to lunches, and later he would refrain from driving and refused rides as well. Without Simon, the regularity of the lunches too waned away.

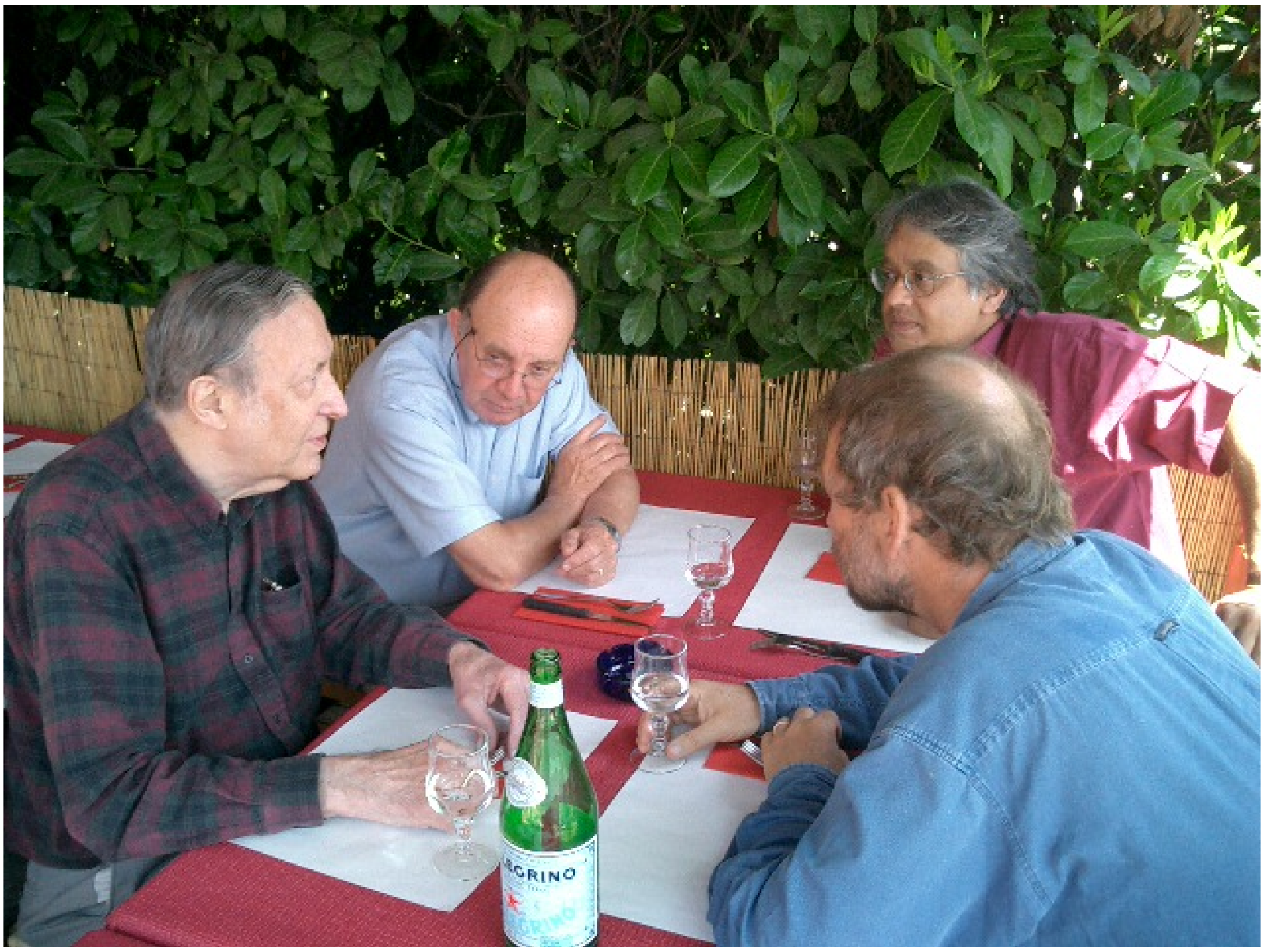

**Fig. 13:** One of the last lunches with Simon, May 2004. Clockwise from left: Simon, Ted Wilson, who too had retired, the present author and Flemming Pedersen.

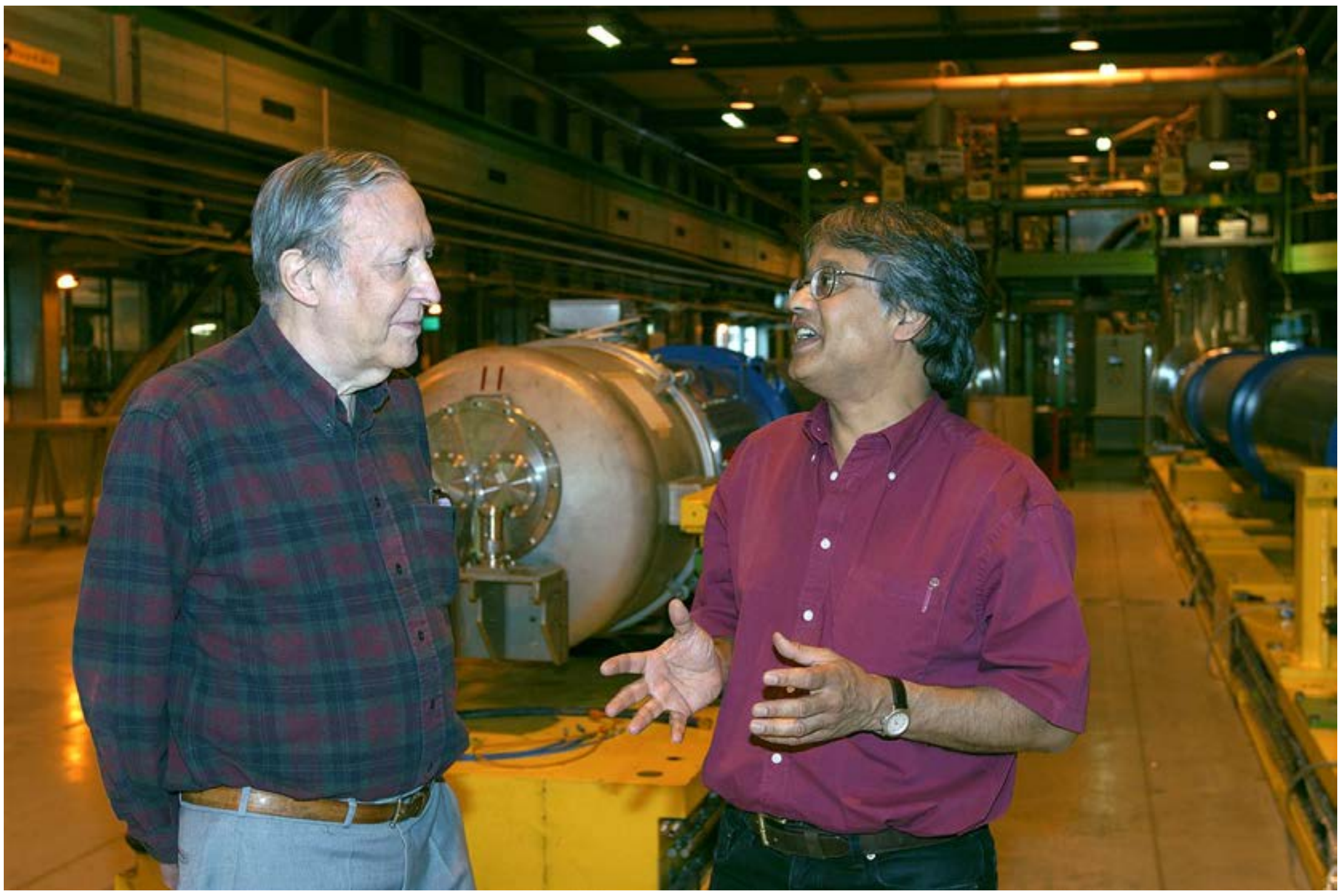

**Fig. 14:** Simon visiting the SM18 LHC Magnets Facility

Simon van der Meer left his mark in many ways in the nearly sixty years of CERN’s existence and of accelerators in general. Although accelerator physicists continue to exude new developments and add to our growing knowledge, Simon’s practicality and general approach provided a dramatic added value not equalled in the history of CERN. It is a moot point whether, in large science projects anywhere, such a personality could still shine out, in the face of the growth of Weberian hierarchies in the name of increased oversight and scrutiny.

To conclude, I can but quote Mathijs van der Meer, in his epitaph for his father in March 2011:

> Simon van der Meer had grown up in a religious family but the war made him lose his faith. Still, he would not have minded a passage from Cardinal Newman’s prayer, “The mission of my life” which says, “I am a link in a chain, a bond of connection between persons”. So even though Simon did not believe in soul or life after death, he would probably agree that there is such a thing as spirit which would remain—in and with the people with whom he interacted.

Among others, I was one of those.

## Acknowledgements

I am grateful to Mathijs and Esther van der Meer for confiding in me some details of Simon’s early period in Holland. Apart from my own records, I have used the CERN document archives, including a *CERN Courier* article (June 2011 issue); I would also like to thank our colleague and friend, Dieter Mohl, who unfortunately is no longer with us since May 2012. I would not have written this article if it had not been for Dieter’s wish and encouragement, which was at the behest of the Editors of *Review of Accelerator Science* (*RAST*), who solicited a personal account for their once-yearly publication to honour Simon van der Meer. A somewhat shorter version of this report has been published in Rev. Accel. Sci. Tech. 4 (2011) 279-291.

## Appendix A: The van der Meer technique for luminosity calibration

Simon's succinct three-page paper ISR-PO/68-31 [9] described a simple and elegant method to evaluate the luminosity delivered by colliding beams at a beam intersection point. This original idea is still widely used by experimental physicists to calibrate their physics detectors at intersection points and give them a measure of the experiment's recorded luminosity. This calibration is usually done at the beginning of a run, and this is still true today at the LHC.

Simon's method required the installation of simple, fast counting devices rather than an intrusive beam-intercepting wire device. The two colliding beams are displaced vertically relative to each other, and the interaction rate is plotted as a function of the displacement. The mean width of the resulting curve gives the effective height of the beams; this value, together with the values of the beam current for each beam, then yields the luminosity. The drawing shown in Fig. 15, reproduced from the 1971 CERN Annual Report, perhaps best describes the simplicity and elegance of Simon's innovation. Of course, much has evolved since then to take additional issues in recent colliding machines into account, such as the crossing angle, the collision offset, and non-Gaussian beam profiles. The original concept is, however, very much in vogue even today at the LHC, accompanied by the usual disparities between the CMS and ATLAS experiments and the ensuing discussions, just as in the days of the UA1 and UA2 experiments at the SPS p–p̄ collider.

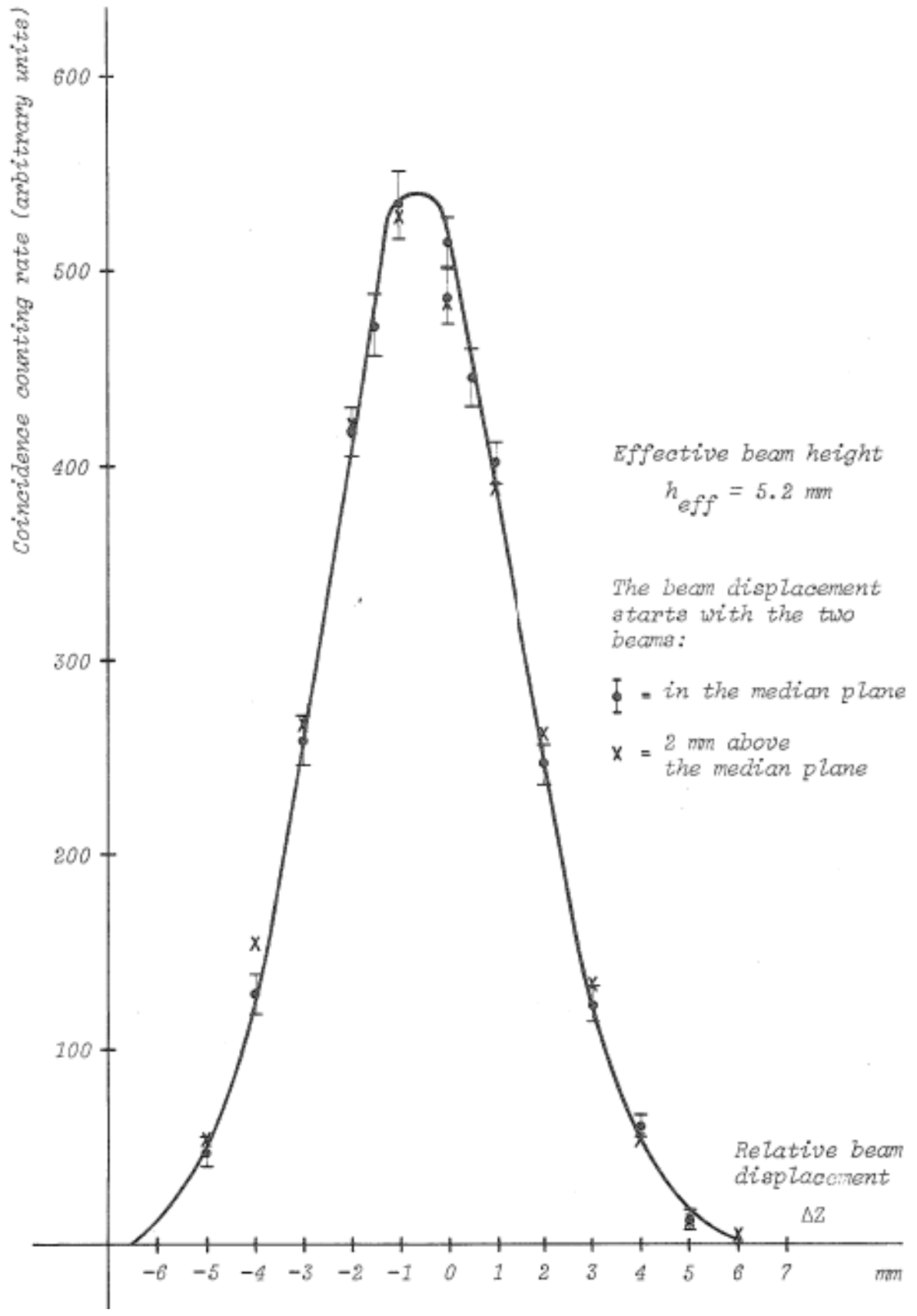

*Figure 8 — Typical measurement to obtain the effective height ($h_{eff}$) of the beams at the intersections. The luminosity can then be obtained for currents of $I_1$ in Ring 1 and $I_2$ in Ring 2 through the formula: $L \simeq 10^{29}(I_1 \cdot I_2)/(h_{eff}\text{ in mm})\ cm^{-2}s^{-1}$.*

**Fig. 15:** reproduced from the CERN Annual Report, 1971